\documentclass[12pt]{article}
\usepackage{epsf}

\newcommand{\be}{\begin{equation}}
\newcommand{\ee}{\end{equation}}
\newcommand{\bmt}{\begin{array}}
\newcommand{\emt}{\end{array}}

\newcommand{\thM}{\tanh{\left(\frac{M}{2T}\right)}}

\newcommand{\intn}[1]{\int{\rm d}^{#1} x}

\newcommand{\lp}{\left(}
\newcommand{\rp}{\right)}
\newcommand{\lc}{\left[}
\newcommand{\rc}{\right]}
\newcommand{\tqed}{${\rm QED_3}$}

\begin{document}

\title{Two-loop corrections to the topological mass term 
in thermal \tqed}
\author{F. T. Brandt$^\dagger$, Ashok Das$^\ddagger$,
J. Frenkel$^\dagger$ and K. Rao$^*$ 
\\ \\
$^\dagger$Instituto de F\'\i sica,
Universidade de S\~ao Paulo\\
S\~ao Paulo, SP 05315-970, BRAZIL\\
$^\ddagger$Department of Physics and Astronomy\\
University of Rochester\\
Rochester, NY 14627-0171, USA\\
$^*$Department of Physics, University of Connecticut\\
Storrs, CT 06269, USA}
\maketitle

\bigskip
\noindent

\begin{abstract}
We study the radiative corrections to the Chern-Simons mass term at
two loops in $2+1$ dimensional quantum electrodynamics at finite
temperature.  We show that, in contrast to the behavior at zero
temperature, thermal effects lead to a non vanishing contribution 
at this order. Using this result, as well as the  large gauge
Ward identity for the leading parity violating terms in the static 
limit, we determine the leading order parity violating effective
action in this limit at two loops, which generalizes the one-loop
effective action proposed earlier. 
\end{abstract}
\vfill\eject

It is well known that, in odd space-time dimensions, one can
add to the usual Maxwell term a parity breaking Lagrangian of the gauge field,
known as the Chern-Simons (CS) term \cite{CS}. In three dimensional QED,
for example, the CS action has the form \cite{DJT,H}
\be\label{1}
S_{\rm CS} = \frac m 2 \int d^{3}x\, \epsilon^{\mu\nu\lambda}
\left(\partial_{\mu}A_{\nu}\right)A_\lambda .
\ee
This action, which is invariant under gauge transformations,
provides a tree level mass $m$ for the gauge field. When fermions are
interacting with the gauge field, the coefficient of the CS term can receive
corrections through fermion loops. Thus, for example, the one-loop
correction, at zero temperature, shifts the value of the tree level
mass as \cite{Redlich} 
\be\label{2}
m\rightarrow m + \frac{e^2}{4\pi}, 
\ee
where we have assumed, for simplicity, a single flavor of 
fermion with charge $e$ and mass $M>0$, interacting with the Abelian
gauge field. In an interesting paper \cite{coleman}, Coleman and Hill
have  shown that, at zero temperature, higher order corrections to the
CS mass term are absent in  \tqed. The proof  essentially uses
two key properties of the amplitudes of the underlying theory, 
namely,
\noindent
i) the Abelian gauge invariance of the theory and
\noindent
ii) the analytic behavior of the amplitudes at zero temperature.

On the other hand,  while gauge invariance still holds at finite
temperature, thermal amplitudes violate analyticity. This happens because
new branch cuts develop at finite temperature, as a result of
possible additional channels of
reaction, so that the thermal amplitudes are non-analytic at the origin
in the energy-momentum plane. For example, at nonzero temperature, the
one-loop radiative correction due to a single flavor of fermion, in
the static limit, shifts the tree level CS mass as \cite{babu,Kao,Zuk}
\be\label{3}
m\rightarrow m + \frac{e^2}{4\pi}\thM,
\ee
while a very different behavior is observed in the long wavelength
limit \cite{BDF1}. Since analyticity, which is a key ingredient in the
proof given by Coleman and Hill, is violated by thermal amplitudes,
one would expect that higher order corrections to the CS term may be
non vanishing at finite temperature.

In this note, we show explicitly that the CS term indeed receives
corrections at two loops in thermal \tqed.
Such a calculation is, in general, quite involved and cannot be
performed in closed form. A great simplification, however, occurs at high
temperature, in the static limit, where we find that the leading
correction to the coefficient of the CS term, at two loops, is given by
\be\label{4}
\Pi_{2}^{(2)}(T\gg M,m)\simeq\lp 2\, m - 3\, M\rp \frac{e^4}{192\,\pi^2\,T^2}
\ln\lp\frac T m\rp
\ee
In fact, there is every reason to believe that, at finite temperature,
the CS term will, in general, receive nonzero corrections at all higher loops.

Since the CS mass term gets radiative corrections at higher orders, one
can ask how the structure of the effective theory will modify so as to
be  invariant under large gauge transformations. In this connection,
let us recall
that, at finite temperature, the time direction becomes compact so
that nontrivial large gauge transformations, associated with this
topological feature, can arise even in an Abelian CS theory
\cite{DLL,DD}. As was shown in \cite{BDF1}, it is really in the
static limit that the question of large gauge invariance comes up. In this
limit of thermal \tqed, the leading order large gauge invariant,
parity violating effective  action, which results from  the one-loop
fermion  contributions, coincides with the exact one-loop parity
violating effective action 
evaluated in the special background $A_0=A_0(t)$, $\vec A=\vec A(\vec
x)$ and has the form
\cite{DGS,FRS}
\be\label{5}
\Gamma_{\rm PV}^{(1)}=\frac{e}{2\pi}\intn{2}
\arctan\lc\thM\tan\lp\frac{e\,a}{2}\rp\rc\,B,
\ee
where $a=\int_0^{1/T}{\rm d} t A_0\lp t\rp$ and 
$B=\epsilon^{0ij}\partial_i A_j$ is the magnetic field. Note that, for
an even number of fermion flavors,
Eq. (\ref{5}) is invariant under the large gauge
transformations $ea\rightarrow ea + 2\pi n$, where $n$ is an integer
(with the magnetic flux quantized).

Since the two-loop contribution to the CS term is non-zero at
finite temperature, we would expect all parity violating amplitudes to
receive non vanishing two-loop corrections as well, if large gauge
invariance  were to hold. In fact, one can ask how the leading order
effective action in the static 
limit in Eq. (\ref{5}) would modify at two loops to have manifest
large gauge invariance. This can be determined systematically with the
help  of the
large gauge Ward identity \cite{DDF}, which reflects the large gauge
invariance of the theory in the static limit. The solution of this
Ward  identity
shows that the form of the leading order effective action, at two loops, is
similar to that in Eq. (\ref{5}), with the simple replacement 
\be\label{6}
\thM\rightarrow\thM + \frac{4\pi}{e^2}\Pi_2^{(2)}\lp T\rp.
\ee

We describe next only the main steps of our analysis. The correction
to the CS mass term, at two-loops, 
can be obtained from the photon self-energy diagrams shown in Fig 1.
\begin{figure*}
    \epsfbox{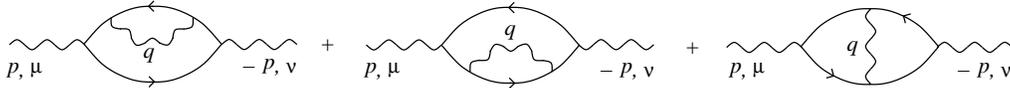}
\bigskip
\caption{Second order contributions to the vacuum polarization.} \label{fig1}
  \end{figure*}
In thermal \tqed, the vacuum polarization continues to be transverse
to the external momentum (as at zero temperature) and has the form
\be\label{7}
\Pi^{\mu\nu}(p,u)=\Pi_1^{\mu\nu}(p,u) + i\epsilon^{\mu\nu\lambda}
p_\lambda \Pi_2(p,u),
\ee
where $\Pi_1^{\mu\nu}$ denotes the parity conserving part of the
photon self-energy, which is transverse and symmetric in the Lorentz
indices (At finite temperature, this structure is a combination of two
independent transverse structures.). Here, we are interested only in
the  parity violating part
of the self-energy, which is given by the second 
term in Eq. (\ref{7}). Note that these functions depend, in general,
on the velocity $u^\mu$ of the heat bath. The coefficient of the CS term,
$\Pi_2(0,u)$, can be projected out from $\Pi^{\mu\nu}$ as
\be\label{8}
\Pi_2(0,u)=\frac{1}{2i}\epsilon_{\mu\nu\rho}
\lc\frac{p^\rho}{p^2}\Pi^{\mu\nu}(p,u)\rc_{p=0}.
\ee

In order to compute the two-loop vacuum polarization at finite
temperature, it is convenient to consider first the fermion box
diagrams $B^{\mu\nu\alpha\beta}(p,q,u)$ depicted in Fig. 2.
The vacuum polarization can then be obtained from these by attaching
together the
photon lines with momenta $q$ and $-q$, giving rise to a  photon
propagator $D_{\alpha\beta}(q)$, which, at zero temperature, has the form
\be\label{9}
D_{\alpha\beta}(q)=\frac{1}{q^2-m^2}\lp\eta_{\alpha\beta}-
\frac{q_\alpha q_\beta}{q^2} - i\,m\,\epsilon_{\alpha\beta\sigma}
\frac{q^\sigma}{q^2}\rp + \xi \frac{q_\alpha q_\beta}{q^4},
\ee
Here $\xi$ is the gauge fixing parameter and the photon propagator
contains a parity breaking term coming from the CS action given in
Eq. (\ref{1}).

\begin{figure*}
    \epsfbox{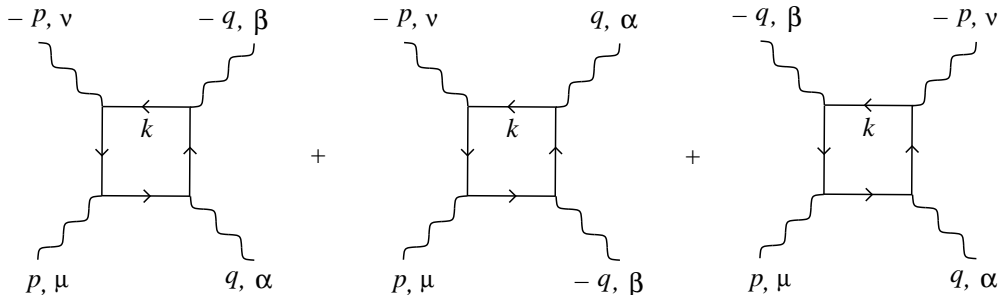}
\bigskip
\caption{Box diagrams which contribute to the photon self-energy at
  two-loop order.} \label{fig2}
  \end{figure*}

The other two photon lines in Fig. 2, carrying momenta $p$ and $-p$,
become the external lines of the self-energy $\Pi^{\mu\nu}(p)$. Thus, we
see that the evaluation of the two loop CS coefficient in Eq. (\ref{8})
requires the study of the quantity
\be\label{10}
B(q)=\epsilon_{\mu\nu\rho}\lc\frac{p^\rho}{p^2}
B^{\mu\nu\alpha\beta}(p,q,u)\rc_{p=0}D_{\alpha\beta}(q).
\ee
As a result of the gauge invariance of the box amplitude, namely,
\be\label{11}
q_\alpha\,B^{\mu\nu\alpha\beta}(p,q,u) = 0,
\ee
it follows that the terms proportional to $q_\alpha$ in the photon
propagator (\ref{9}) do not contribute to Eq. (\ref{10}), so that
$B(q)$ is effectively a gauge invariant quantity (which is not true in
a non-Abelian theory). It is important to note that
only the linear terms in $p$ from the box diagrams can possibly
contribute to $B(q)$, since this quantity is evaluated in the limit
$p\rightarrow 0$. This fact has the interesting consequence that
$B(q)$, as defined in Eq. (\ref{10}), vanishes at zero 
temperature \cite{bl}. 
This happens because the box amplitude is an analytic
function at $T=0$, which, together with Eq. (\ref{11}), actually implies that
$B^{\mu\nu\alpha\beta}(p,q)$ must be at least quadratic in 
the external momentum $p$ \cite{coleman,BDF1} (and, consequently,
$B(q)$ must vanish in the limit $p\rightarrow 0$). However, this is not true
at finite temperature, since in this case the amplitudes are no longer
analytic functions. In fact, as shown in \cite{BDF1}, the thermal box
graphs do give leading order contributions which are linear in $p$.

In order to compute the thermal effects, we will employ the
imaginary-time formalism \cite{kapusta,lebellac,das}, where the
integration over continuous energies is replaced 
by a summation over the discrete values $\pi\,i\,n\,T$, where $n$ is an
odd or even integer respectively for fermions or bosons. 
Using the relations (\ref{9}), (\ref{10}) and  (\ref{11}), the
coefficient of the CS term, given by Eq. (\ref{8}), can be written, at
two loops, as
\begin{eqnarray}\label{12}
&\Pi_2^{(2)}(0,u)\equiv\Pi_2^{(2)}(T)=\displaystyle{\frac{T}{2i}
\sum_{q_0=2\pi\,i\,n T}\int\frac{{\rm d}^2 q}{(2\pi)^2} B(q)}=\nonumber\\
&\displaystyle{\frac{T}{2i}\lim_{p\rightarrow 0}\lc
\epsilon_{\mu\nu\rho} \frac{p^\rho}{p^2} 
\sum_{q_0=2\pi\,i\,n T}\int\frac{{\rm d}^2 q}{(2\pi)^2} 
\lp \eta_{\alpha\beta}
-i\,m\, \epsilon_{\alpha\beta\sigma} \frac{q^\sigma}{q^2}\rp
\frac{B^{\mu\nu\alpha\beta}(p,q,u)}{q^2-m^2}  \rc}, 
\end{eqnarray}
where we have emphasized the temperature dependence of the CS 
coefficient in $\Pi_2^{(2)}(T)$. 

To evaluate explicitly the above expression for $\Pi_2^{(2)}(T)$, we
first compute
the trace over the Dirac matrices in $B^{\mu\nu\alpha\beta}$ and
Taylor expand the result up to linear terms in the external momentum
$p$. The terms of the resulting expression are either proportional to
an $\epsilon$ tensor multiplied by odd powers of the fermion mass $M$, or
contain even powers of $M$ with no  $\epsilon$.
As a consequence of the Bose symmetry of $B^{\mu\nu\alpha\beta}$
(and taking into account the fact that terms which are odd in the
internal photon
and fermion momenta vanish by symmetric integration) the contractions
with the $\eta_{\alpha\beta}$ and $\epsilon_{\alpha\beta\sigma}$ tensors in
Eq. (\ref{12}) give a non-vanishing result only for the terms
involving odd and even powers of $M$, respectively.

There are two meaningful limits of $p\rightarrow 0$ that can be considered in
Eq. (\ref{12}), namely, the static and the long wavelength limits.
In this note, we discuss only the static limit 
$\vec p\rightarrow 0, p_0 = 0$, which is
simpler than the long wavelength limit $p_0\rightarrow 0,\vec
p=0$. The reason is that no analytic continuation for the external
energy is necessary in the static limit, so that the linearization 
in the external
momentum in the box diagram can be performed before the computation 
of the sum over the thermal energies. Furthermore, as we have already
shown \cite{BDF1}, the long wavelength limit is manifestly large gauge
invariant. 

The next step consists in using the Feynman parameterization to
integrate over the fermion momentum $\vec k$. The integration over the
photon momentum $\vec q$ can also be done using this method. We must
finally  perform, in addition to the summation
over $q_0$, a further summation over the thermal energies 
$k_0=(2\, l+1)\pi\,i\,T$ of the fermion, where $l$ is an
integer. Unfortunately, it is not possible to obtain a closed form
expression for these sums in general. But an important simplification
occurs at high temperatures, $T\gg M,m$, where  the
leading thermal correction is obtained from the zero mode
$q_0=0$. After performing the integration over the Feynman parameter, 
we find that
\begin{eqnarray}\label{13}
\Pi_2^{(2)}(T\gg M,m) & \simeq & 
\displaystyle{\lp 2\,m - 3\,M\rp \frac{e^4}{4\pi^2}
T^2\,\ln\lp\frac T m\rp \sum_{k_0}\frac{1}{k_0^4}} \nonumber \\
& = & 
\displaystyle{\lp 2\,m  - 3\,M \rp \frac{e^4}{192\pi^2\,T^2} 
\ln\lp\frac T m\rp}.
\end{eqnarray}
It is interesting to remark that the above expression is well behaved,
thanks to the tree level CS mass $m$, which provides an infrared
cutoff. (Note, however, that for $M\neq 0$, this expression is
logarithmically divergent as $m\rightarrow 0$, a feature special to the
thermal field theory.) 

Since the two-loop parity violating amplitudes are non-vanishing at
finite temperature, we can inquire about the structure of the large
gauge invariant, parity violating  
effective action at two loops. This can be done following the
analysis in \cite{BDF1}. Let us assume that the leading term of the full
parity violating effective action (in the static limit) can be written as
\be\label{14}
\Gamma_{\rm PV}=\frac{e}{2\pi}\intn{2}\tilde\Gamma(\tilde a) B,
\ee
where $\tilde a=e\, a = e\,\int_0^{1/T}{\rm d} t A_0(t)$. We can now
derive, as in \cite{BDF1}, a large gauge Ward identity that this
effective  action must satisfy for large gauge invariance to
hold. In fact, it can be easily checked  that the quantity 
$\tilde\Gamma(\tilde a)$ introduced in Eq. (\ref{14}) has to satisfy 
\be\label{15}
\frac{\partial^2\tilde\Gamma}{\partial\tilde a^2}=
\frac 1 2\lc\frac{1}{2\tilde\Gamma^\prime(0)} - 
2\tilde\Gamma^\prime(0)\rc \frac{\partial\tilde\Gamma}
{\partial\tilde a}\sin(2\tilde\Gamma),
\ee
where $\tilde\Gamma^\prime(0)=\lp\partial\tilde\Gamma/\partial\tilde
a\rp_{\tilde a = 0}$ 
denotes the one-point function. This nonlinear Ward identity, which
relates the amplitudes obtained in perturbation theory, shows that all
the parity violating amplitudes are connected recursively to the
one-point  function. The solution of Eq. (\ref{15}) is easily seen to be
\be\label{16}
\tilde\Gamma(\tilde a)=\arctan\lc 2\tilde\Gamma^\prime(0)
\tan\lp\frac{\tilde a}{2}\rp\rc,
\ee
where $\tilde\Gamma^\prime(0)$, up to two-loop order is given by
\be\label{17}
\tilde\Gamma^\prime(0)=\frac 1 2 \thM + \frac{2\pi}{e^2}\Pi_2^{(2)}(T).
\ee
Substituting the above expression for $\tilde\Gamma(\tilde a)$ 
into Eq. (\ref{14}), we obtain a result which extends, to two-loop
order, the parity breaking effective action proposed earlier
\cite{DGS,FRS}. (In fact, as we have shown in \cite{BDF1}, the leading
term in the static limit coincides with the effective action in the
special gauge background  $A_0 = A_0 (t)$ and $\vec{A} =
\vec{A}(\vec{x})$.) Note also that if we know
$\tilde{\Gamma}^\prime(0)$ up to any loop (namely, if we calculate the
coefficient of the CS term to any loop), we can obtain the leading
term in the large gauge invariant, parity violating effective action
(in the static limit) up to that order from Eqs. (\ref{16}) and (\ref{14}).

In conclusion, we have shown that, at finite temperature, the CS mass
term, in the Abelian theory, receives a non vanishing radiative
correction  at two
loop order and we expect this to be true at all higher orders. Furthermore,
using the large gauge Ward identity, we have determined the full
parity violating effective action, at two loops, which shows that all
the other parity violating amplitudes must modify as well, in a well
defined manner, for large gauge invariance to hold. 

This work was supported in part by U.S. Dept. Energy Grant DE-FG
02-91ER40685, DE-FG02-92ER40716.00, NSF-INT-9602559 as well as by CNPq, Brazil.

\end{document}